\documentclass[11pt]{article}
\usepackage[margin=1in]{geometry}
\usepackage{graphicx}
\usepackage{booktabs}
\usepackage{amsmath}
\usepackage[hidelinks]{hyperref}
\usepackage{xcolor}

\title{Personal Salience: Highlighting Is Social, but\\
Individuality Lives in Selection}
\author{
  Kazuki Nakayashiki \quad Keisuke Watanabe \\[3pt]
  Glasp Inc. \\
  \texttt{kazuki@glasp.co} \quad \texttt{kei@glasp.co} \\[3pt]
  {\small\itshape Co-first authors (equal contribution).}
}
\date{}

\begin{document}
\maketitle

\begin{abstract}
Social highlighters let people mark the passages that matter to them as they
read. We ask how much of an individual is recoverable from these naturalistic
highlighting traces, using a co-readership identity control --- the same document
highlighted by many users --- that holds the document and topic fixed and asks
whether a person's own history predicts their marks better than another reader's
does. We separate generic salience (document structure), crowd salience (what
other readers marked), and personal salience (the individual residual). Two
findings dominate. First, \emph{highlighting is social}: which sentences a person
marks within a document is predicted far better by the crowd than by document
structure or by a personal model, and even a well-estimated crowd --- an
information-privileged baseline that observes other readers' marks on the \emph{same}
document --- beats a frontier LLM ``twin'' conditioned only on the user's
other-document history; the personal signal recoverable in within-document salience
is at most a whisper (own-versus-other identity gap $+0.017$ average precision by the
embedding scorer, small but significant; $+0.012$, n.s., by the twin). Second, and in sharp contrast,
\emph{individuality lives in selection}: when the task is choosing which of the
already-salient passages are yours, a person's own history is a strong, leakage-free
predictor (own-versus-other identity gap $+0.14$; popularity is in fact
\emph{anti}-predictive here, scoring below chance, because the most-marked spans are
everyone's). A topic decomposition shows this selection signal is \emph{largely
stable thematic preference} --- it shrinks $\sim$8$\times$ against a topically-matched
peer; a thin residual remains even against a near-topic-twin, but we cannot tell
whether it is non-thematic ``style'' or finer-grained topic. The non-obvious part is
an \emph{asymmetry}: under the same embedding scorer the individual signal is about
$6$--$8\times$ weaker (nearly an order of magnitude) in within-document salience
($+0.017$) than in selection ($+0.10$ to $+0.14$). For users with rich histories the selection
signal is captured by a few dozen highlights, after which it plateaus. We also report a suggestive population-level effect: \emph{crowd consensus
tends to decline with document popularity} --- the more readers a document has, the less
they agree on what to highlight, even at a fixed crowd size. Methodologically, we
show that naive history-conditioning evaluations leak (the target document's own
highlights enter the profile in $\sim$42\% of pairs, inflating personal scores by
$+0.07$ to $+0.15$ AP) and that capping the crowd at a few readers
overstates personalization; our results are leakage-free, use a dense crowd, and
employ a model-matched identity control. Highlights carry a genuine individual
signature, but it is a thin layer over a strong shared one, and it surfaces far more
in which salient things a person selects than in what is salient.
\end{abstract}

\section{Introduction}
A common intuition, and a recent research program around digital twins and
generative agents~\cite{park2024,santurkar2023}, holds that a person's behavioral
traces reveal their interests and predict their choices. Highlighting --- marking
the passages that resonate while reading --- is an unusually rich, low-cost,
in-the-wild trace. We ask how much of an individual is actually recoverable from
it.\footnote{Code and reproducibility bundle (the cluster-bootstrap estimator, the
paper, and figures): \url{https://github.com/glasp-co/personal-salience}}

The core question is whether what a person highlights is a personal fingerprint,
recoverable from their own past behavior, or mostly shared salience that any reader
would mark. We separate three quantities that personalization work usually
conflates: \textbf{generic salience} (semantic centrality, a content-only proxy for
structural importance), \textbf{crowd salience} (what other readers actually marked),
and \textbf{personal salience} (the individual residual). The separation is possible because Glasp
materializes co-readership: for documents highlighted by multiple users we know
exactly who marked what, giving a within-document identity control rarely available
in personalization research. We study two tasks: predicting which \emph{sentences}
of a document a person highlights (salience), and predicting which of the
\emph{already-highlighted passages} are a given person's (selection).

\paragraph{Contributions.}
(1) A clean decomposition of highlighting into generic, crowd, and personal
components, with a co-readership identity control. (2) Evidence that within-document
highlighting is \emph{social}: the crowd --- though an information-privileged
baseline whose edge over the twin is expected, not surprising --- dominates both
generic structure and the user's own history, and the within-document personal
residual is at most a whisper (embedding gap $+0.017$). (3) The
contrasting finding that individuality lives in \emph{selection}: a leakage-free
own-versus-other personal signal (identity gap $+0.14$), measured on users with
sufficient history, which a topic decomposition shows is largely stable thematic
preference --- with a thin residual we cannot separate from finer-grained topic ---
and which, under the same scorer, is about $6$--$8\times$ stronger in
selection than in salience. (4) A methodological caution: history-conditioning
evaluations leak, and small crowds overstate personalization; we quantify both and
correct for them. Separately --- as an exploratory observation, not a headline claim
--- we note that crowd consensus may decline with document popularity (the trend is
not statistically conclusive; \S\ref{sec:decay}).

\section{Related work}
\textbf{Highlighting as shared salience.} Aggregated highlights (e.g.\ Kindle
Popular Highlights) act as collective reader response, and human highlights predict
comprehension and interest and correlate with summaries~\cite{winchell2020,cho2020}.
We quantify this predictively and contrast it with the personal component.
\textbf{Social annotation and collaborative tagging.} Systems for shared, in-place
annotation~\cite{zyto2012} and collaborative tagging exhibit stable, largely shared
structure tied to imitation and common knowledge~\cite{golder2006}. That work studies
the collective artifact and its dynamics; our novelty is to use co-readership of the
\emph{same} document as a leakage-free identity control that separates the individual
from the crowd, rather than modelling the aggregate.
\textbf{Annotation disagreement.} Crowdsourcing and perspectivist work show that
disagreement is signal, not noise, and that larger, more diverse annotator pools
surface lower agreement~\cite{aroyo2015,plank2022}; our (suggestive) popularity-linked
consensus decline is a naturalistic, fixed-crowd-size instance of this. \textbf{Wisdom of
crowds.} Aggregates are strong because diverse errors cancel~\cite{surowiecki2004};
we show the crowd is a strong predictor of an \emph{individual} precisely because
highlighting is largely shared, yet its predictive power erodes as the audience
diversifies. \textbf{Personalization and individual differences.} Reading and
annotation are idiosyncratic~\cite{shardlow2022}; personalized saliency exists in
vision~\cite{xu2017} and video highlights~\cite{gygli2018}; conditioning LLMs on a
user's history is the basis of personalization benchmarks like
LaMP~\cite{salemi2023}, and steering by demographics is insufficient for individual
opinions~\cite{santurkar2023}. That personalized saliency is recoverable in vision
might predict the same for text, but we find the opposite \emph{within} a document:
textual salience is far more consensual --- the crowd baseline is much stronger than
any image-saliency prior --- and the recoverable individual signal instead surfaces
in selection. Our contribution is the within-document identity control, the
leakage-free measurement, and the finding that the recoverable individual signal
lives in selection rather than salience.

\section{Data}
Glasp is a web highlighter with hundreds of thousands of active users and millions
of highlighted URLs. We use only public-web articles and exclude Kindle (copyright)
and PDF (frequently sensitive). Co-readership is materialized: a large pool of
documents are highlighted by two or more users, and for such a document we can
enumerate the co-readers and the exact spans each one marked. Documents per user
are heavy-tailed; the corpus skews to English web articles. All analysis is
read-only on internal data; the per-pair results derive from user highlighting
behavior and are not released. We report aggregate effect sizes with confidence
intervals, and the estimator is provided as code.

\section{Method}
\paragraph{Tasks.}
\emph{Within-document task.} We fetch the article body, segment it into sentences,
and re-anchor each co-reader's stored spans to sentences (a drift gate keeps a user
only if at least half of their spans re-anchor). The candidate set is all
sentences; positives are the sentences a user highlighted.
\emph{Selectivity task.} The candidate set is the union of the spans the
co-readers highlighted; for each co-reader we predict which of these
already-salient spans are theirs.

\paragraph{Scores.}
\textbf{Crowd}: how many \emph{other} co-readers marked a candidate. Crucially we
estimate it from a \emph{dense} crowd (all available co-readers, median 38), not the
$\leq$5 of earlier pilots. \textbf{Generic}: sentence centrality (cosine to the
document embedding centroid). \textbf{Personal M1}: cosine to a mean-pooled
embedding of the user's past highlights. \textbf{Twin}: a proprietary LLM (gpt-5.5)
shown up to 50 of the user's past highlights and asked to rank the candidates; the
model is closed, so this baseline is not independently reproducible and we treat it
as one extractor among several rather than as a ceiling on what a personalizer could
do. \textbf{Other} (the identity control): the \emph{same} method applied with
another co-reader's profile --- the \emph{most prolific} co-reader, a deliberately
conservative choice that gives the control the most data and thus the best chance to
match. The identity control is method-matched (twin-own vs.\ twin-other; M1-own vs.\
M1-other), so a positive own-minus-other gap cannot be attributed to the extractor's
general skill.

\paragraph{Leakage-free profiles.}
The personal profile is built \emph{per pair}, excluding the exact target document,
so a user's own marks on the target never enter their profile. (A single cached
exclusion, as in naive pipelines, leaks the target in $\sim$42\% of within-document
and $\sim$31\% of selectivity pairs; we measure the inflation directly.)

\paragraph{Cohorts.}
Co-readership filters differ by task, so the cohorts are not identical: the
within-document task keeps documents with $\geq$10 readers (1{,}004 pairs);
selectivity requires a non-trivial union of co-reader spans (493 pairs, median 34
candidates per pair); and the scaling-law analysis further restricts to users with
$\geq$100 prior highlights so that $K$ can be swept to 100 (476 pairs) --- a
power-user subset whose absolute levels (plateau $\sim$$0.335$) therefore differ
from the full selectivity cohort ($0.397$). Every contrast is computed within a
single cohort.

\paragraph{Metrics.}
We rank candidates by each score and report average precision (AP) per (user,
document) pair, with a cluster bootstrap (resampling whole clusters) for
confidence; contrasts are confirmed clustering by document and, separately, by
user. Embeddings use \texttt{text-embedding-3-small} (512d).

\section{Results}

\subsection{Highlighting is social: the crowd dominates within-document salience}
\label{sec:social}
\begin{figure}[t]
\centering
\includegraphics[width=0.7\linewidth]{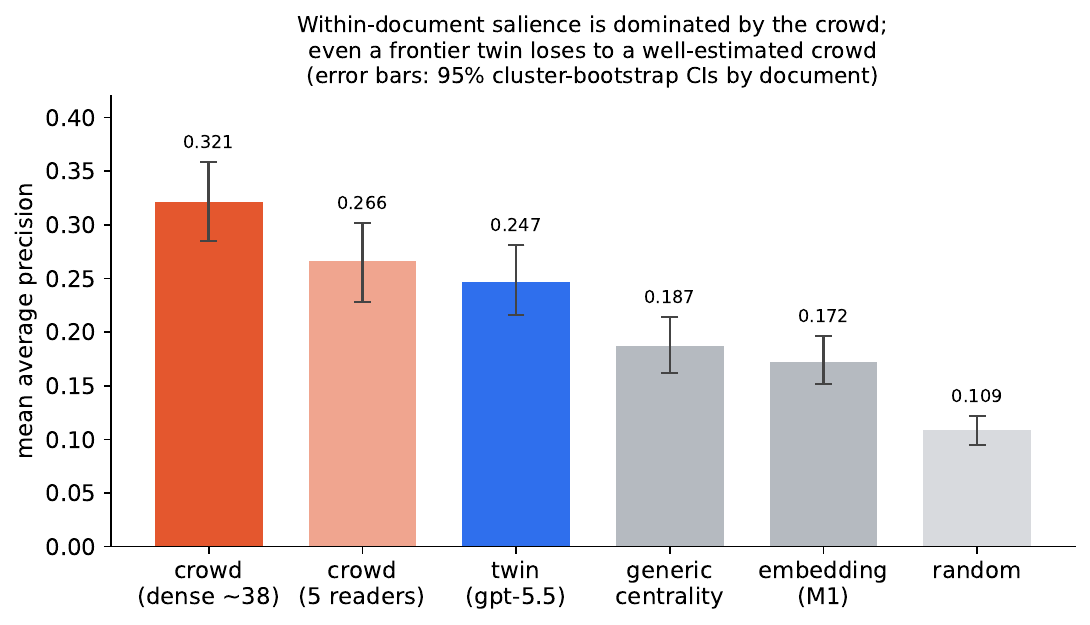}
\caption{Within-document salience is dominated by the crowd; even a frontier twin
loses to a well-estimated (dense) crowd, and the 5-reader crowd of naive pipelines
is artificially weak. Error bars are 95\% cluster-bootstrap CIs (by document); the
crowd's lead is established by \emph{paired} contrasts (crowd minus twin $+0.074$,
CI $[0.047, 0.102]$; crowd minus 5-reader $+0.054$, CI $[0.038, 0.070]$), which are
significant even though the marginal intervals overlap.}
\label{fig:within}
\end{figure}

On the within-document task (1{,}004 pairs, 848 users, dense crowd median 38;
Table~\ref{tab:within}, Figure~\ref{fig:within}), the crowd is the strongest
predictor by a wide margin: crowd $0.321$ versus generic centrality $0.187$, a
mean-pooled embedding profile $0.172$, and the frontier twin $0.247$. What predicts
which sentences you highlight is overwhelmingly \emph{what other readers
highlighted}, not document structure, and not your own history. Even the frontier
LLM twin loses to a well-estimated crowd ($-0.074$); it loses even to a 5-reader
crowd ($-0.019$). The artificially small ($\leq$5-reader) crowd of the pilot stage
($0.266$) understated the crowd by $+0.055$ relative to a dense crowd, which is the
size of the apparent ``personalization'' edge. We stress that the crowd is
\emph{information-privileged}: it observes other humans' marks on the exact target
document --- effectively a leave-one-out of the same labels the twin must predict
cold --- so its win over the twin is \emph{expected} and is not a claim that
aggregation out-reasons the model. The substantive finding is that within-document
highlighting is highly consensual: which sentences you mark is well predicted by what
peers marked, and barely improved by knowing whose marks they are.

\begin{table}[t]
\centering
\caption{Within-document task, mean AP (1{,}004 pairs, 848 users, leakage-free,
dense crowd). The crowd dominates; the personal model barely exceeds another
reader's (embedding identity gap $+0.017$, a tiny whisper).}
\label{tab:within}
\begin{tabular}{lc}
\toprule
Scorer & Mean AP \\
\midrule
Crowd salience (dense, $\sim$38 readers) & 0.321 \\
Crowd salience (5 readers)               & 0.266 \\
Twin (gpt-5.5, own profile)              & 0.247 \\
Another reader's twin                    & 0.235 \\
Generic centrality                       & 0.187 \\
Embedding profile (M1, own)              & 0.172 \\
Embedding profile (M1, another reader)   & 0.155 \\
Random base rate                         & 0.109 \\
\bottomrule
\end{tabular}
\end{table}

The recoverable personal signal here is, at most, a \emph{whisper}: the
own-versus-other identity gap is $+0.017$ for the embedding (95\% CI $[0.008, 0.027]$,
small but significant) and $+0.012$ for the frontier twin (95\% CI $[-0.004, 0.029]$,
n.s.). So a tiny within-document personal signal is detectable by the cleaner
embedding scorer though not by the noisier twin --- but either way it is about
$6$--$8\times$ below the selectivity signal ($+0.10$ to $+0.14$, \S\ref{sec:topic}),
and the crowd, not the individual, is what dominates here. On
documents where the crowd is weak (low consensus) the twin does edge it
($+0.043$), but a twin built from \emph{another} reader's profile also beats the
crowd there --- so this reflects the model's general salience judgement, not
personalization (the identity gap stays $\sim$0.02). Individuality does not
``come back'' where consensus breaks down.

\paragraph{Is the crowd just a lead/position prior?} Summarization has a famously
strong \emph{lead} baseline (people mark the opening and the conclusion), so we
checked whether the crowd's dominance is merely positional (a comparable
within-document cohort, 1{,}340 pairs). A pure lead scorer (earlier sentence ranked
higher) reaches AP $0.237$ --- and, strikingly, beats both generic centrality
($0.196$; $+0.041$, 95\% CI $[0.013, 0.074]$) and the personal embedding ($0.185$;
$+0.052$, CI $[0.024, 0.083]$): knowing \emph{which sentences come first} predicts a
person's marks better than knowing their own reading history. Yet the crowd still
beats the lead prior decisively (crowd $0.340$; crowd minus lead $+0.103$, CI
$[0.073, 0.133]$; minus a learned position-decile prior $+0.135$). So the crowd is not
a trivial intro/conclusion effect, and the within-document task is so shared that
position outranks personal history --- reinforcing that individuality is not here.

\subsection{Individuality lives in selection, not salience}
\begin{figure}[t]
\centering
\includegraphics[width=0.92\linewidth]{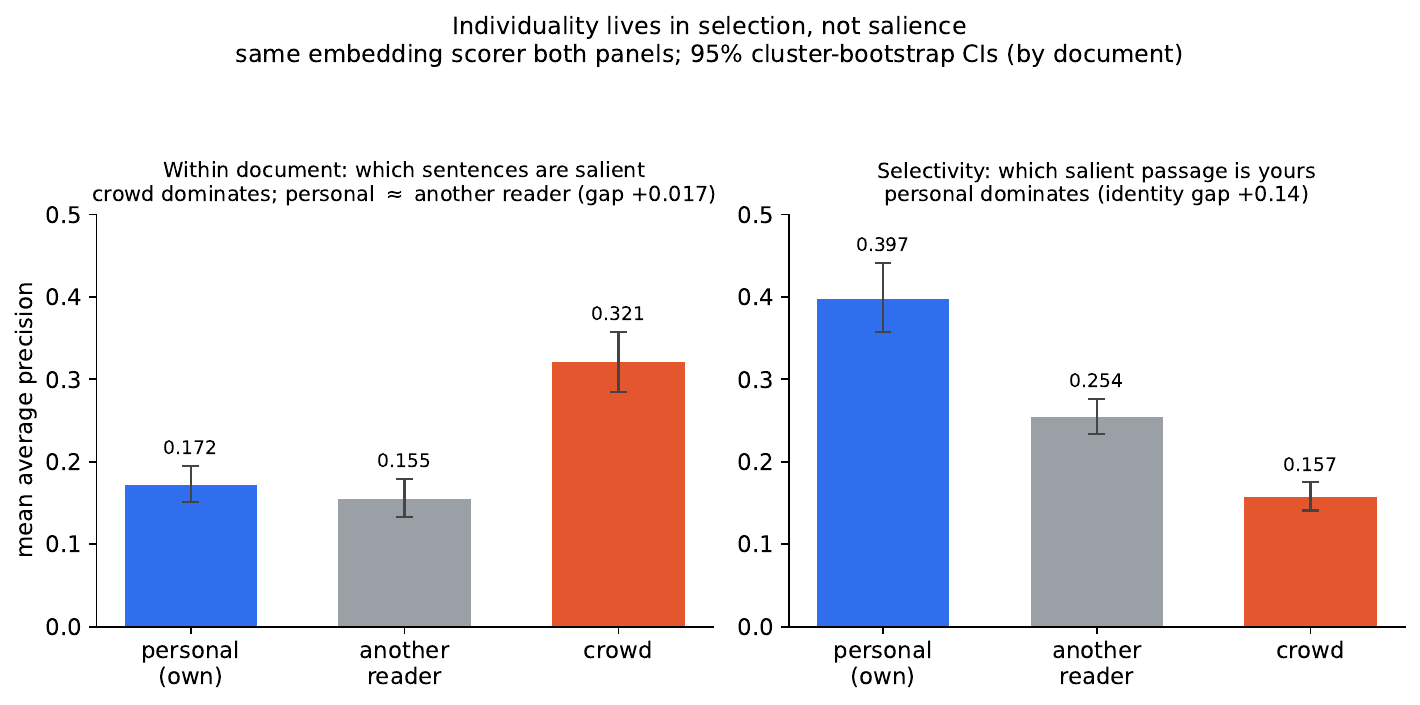}
\caption{Individuality lives in selection, not salience. Both panels use the same
embedding (M1) scorer, so ``personal'' means the same thing in each. Within document
(left) the crowd dominates and the personal model barely exceeds another reader's
(gap $+0.017$, small but significant); in selectivity (right) the personal model
dominates both another reader and the crowd (identity gap $+0.14$). Error bars are
95\% cluster-bootstrap CIs (by document). The two panels use different cohorts
(1{,}004 vs.\ 493 pairs); compare the contrasts \emph{within} each panel, not absolute
heights across panels.}
\label{fig:flip}
\end{figure}

The contrast with selectivity is stark (493 pairs, 462 users;
Figure~\ref{fig:flip}). When the candidates are \emph{already} salient (someone
highlighted them; a median of 34 candidate spans per pair, base rate $0.174$) and the
task is which ones are yours, a person's own history is a strong, leakage-free
predictor: personal $0.397$ versus another reader's $0.254$. The clean,
method-matched contrast is the own-versus-other identity gap, $+0.143$ (positive in
78\% of documents, robust to base-rate filtering, replicated at $+0.135$ on an
independent sample). Crowd popularity is \emph{not} a meaningful competitor here: at
$0.157$ it scores \emph{below} the $0.174$ base rate, because the most-marked spans
are exactly the ones everyone shares and so are least diagnostic of any individual.
We therefore rest the selectivity claim on personal-versus-other, not on the larger
but less meaningful personal-minus-crowd gap. Here the leakage was large ($+0.145$
AP) yet the clean signal survives intact.

The interpretation is principled: the within-document task is dominated by the
variance of \emph{whether a sentence is salient at all} (which is shared), swamping
the individual component; the selectivity task removes that variance by restricting
to already-salient passages, isolating \emph{which} salient thing a person selects
--- and there, individuality is clear.

\begin{table}[t]
\centering
\caption{Own-versus-other identity gaps across tasks, cohorts, and comparison peers
(mean AP difference, 95\% cluster-bootstrap CIs by document). Rows use the embedding
scorer except where noted, so the within-document ($+0.017$) and selectivity gaps are
directly comparable: under the same scorer the individual signal is a tiny but
significant whisper in salience and about $6$--$8\times$ larger in selection.
The selectivity gap shrinks as the comparison peer is matched on topic, leaving a thin
residual we cannot separate from finer-grained topic. The noisier twin does not reach
significance within-document.}
\label{tab:gaps}
\begin{tabular}{lllcc}
\toprule
Task & Cohort & Comparison peer & Gap & 95\% CI \\
\midrule
Within-doc  & main (1{,}004)          & another reader, embedding          & $+0.017$ & $[0.008, 0.027]$ \\
Within-doc  & main (1{,}004)          & another reader, twin               & $+0.012$ & $[-0.004, 0.029]$ (n.s.) \\
Selectivity & main (493)              & another reader, embedding          & $+0.143$ & $[0.105, 0.183]$ \\
Selectivity & $\geq$3-reader (1{,}621) & prolific peer, embedding          & $+0.113$ & $[0.094, 0.133]$ \\
Selectivity & $\geq$3-reader (1{,}621) & topically-nearest, embedding      & $+0.098$ & $[0.081, 0.116]$ \\
Selectivity & $\geq$3-reader (1{,}621) & near-topic-twin (sim$\approx$0.90) & $+0.026$ & $[0.016, 0.037]$ \\
\bottomrule
\end{tabular}
\end{table}

\subsection{What the selection signal is: mostly stable thematic preference}
\label{sec:topic}
Is the selectivity gap a fine-grained ``selection style'', or stable \emph{topic}
preference (you select spans about topics you already like)? To decompose this we
replace the comparison reader with the \emph{topically-nearest} co-reader --- the one
whose leakage-free profile embedding is most similar to the target's --- on a denser
cohort with at least three co-readers (1{,}621 pairs, 1{,}398 users, 142 documents);
Table~\ref{tab:gaps} collects every gap from this section and the last.
A topic-matched peer is a far stronger competitor than the prolific peer (mean
profile similarity rises from $0.61$ to $0.73$) and does predict the target's spans
better (other-AP $0.142\!\to\!0.158$); the gap shrinks accordingly, from $+0.113$
(vs.\ the prolific peer --- this denser $\geq$3-reader cohort's gap is a little below
the main selectivity cohort's $+0.143$, as its larger candidate unions are more
shared) to $+0.098$ (vs.\ the topic-matched peer; 95\% CI $[0.081, 0.116]$, positive
in 100\% of resamples clustering by document and, separately, by user). About 86\% of
the gap survives matching to the average available peer.

\begin{figure}[t]
\centering
\includegraphics[width=0.62\linewidth]{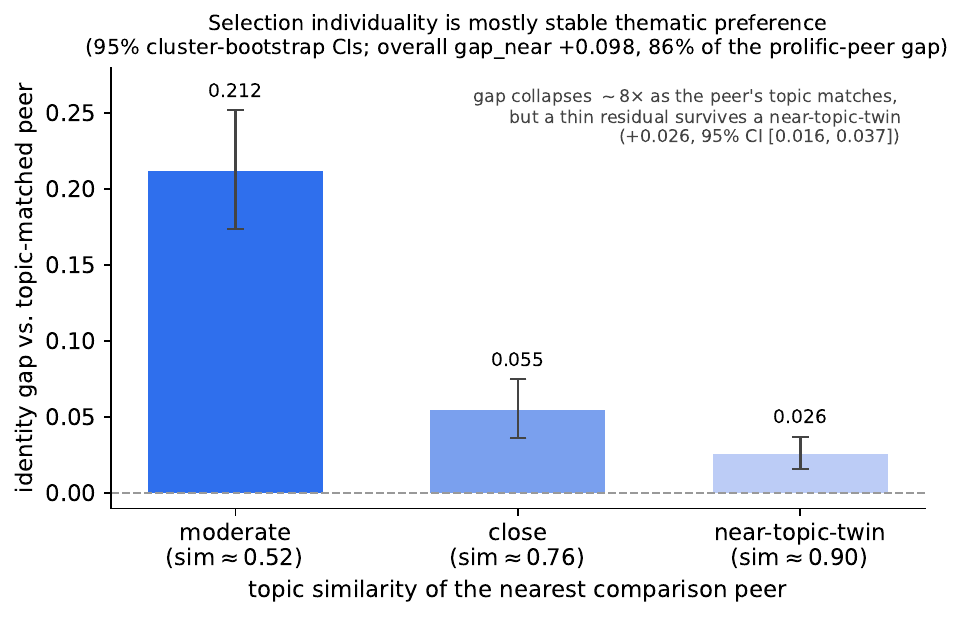}
\caption{Topic decomposition of the selectivity gap. Replacing the comparison reader
with the \emph{topically-nearest} co-reader, the own-vs-other gap collapses
$\sim$8$\times$ as the peer's topic similarity rises (from $+0.21$ to $+0.03$), so the
selection signal is mostly stable thematic preference. A thin residual nonetheless
survives even a near-topic-twin ($+0.026$, 95\% CI $[0.016, 0.037]$; the dashed line
is the ``pure topic'' null). We do not claim the residual is non-thematic style ---
the scorer is semantic and the peer is matched in the same space. Error bars are 95\%
cluster-bootstrap CIs by document.}
\label{fig:topic}
\end{figure}

The decomposition is sharper when we condition on \emph{how} topically close the
nearest peer is. Splitting pairs into tertiles of peer similarity, the gap falls
steeply as the topic match improves: $+0.212$ when the nearest peer is only
moderately similar (sim $\approx 0.52$), $+0.055$ at sim $\approx 0.76$, and $+0.026$
when the nearest peer is a near-topic-twin (sim $\approx 0.90$). The signal is
therefore \emph{strongly} topic-driven --- an $\sim$$8\times$ reduction
(Figure~\ref{fig:topic}) --- so most of the recoverable selection signal is stable
thematic preference. Yet even against a near-topic-twin the residual stays
significantly positive ($+0.026$, 95\% CI $[0.016, 0.037]$, positive in 100\% of
resamples): a thin individual signal survives that coarse shared topic does not
explain.

We are deliberately cautious about what this residual \emph{is}. Our personal scorer
is semantic and the peer is matched in the same embedding space, so the residual may
be finer-grained topic rather than a non-thematic ``style'' (a preference for, say,
conclusions or quantitative claims). Three reasons make $+0.026$ an \emph{upper bound}
on non-thematic style rather than a measurement of it: the nearest peer is only
sim $\approx 0.90$, so $\sim$0.10 of real topic difference still remains; extrapolating
the tertile decay ($0.21\!\to\!0.026$) to a perfect topic twin could land at zero or
at a small positive, and we cannot tell which; and a lightweight embedding
(\texttt{text-embedding-3-small}, 512d) may simply miss fine topic nuance, surfacing
as a spurious residual. The residual is therefore consistent with non-thematic style
being zero. (Note the direction: a \emph{stronger} embedding would capture more topic
and match the peer better, shrinking the residual further --- so a better representation
can only reinforce the ``mostly thematic'' conclusion, not overturn it.) Isolating any
genuine non-thematic component would need structural/positional features or cross-topic
transfer, and is future work. The
honest reading: individuality in selection is \emph{largely stable thematic
preference}. Because the document (and hence its broad topic) is held fixed, this is
preference among the document's \emph{aspects} --- not merely which articles a person
reads --- with a thin residual surviving even topic-matching. The non-obvious part is
the \emph{asymmetry}: under the same embedding scorer this thematic preference is
about $6$--$8\times$ weaker in within-document salience (gap $+0.017$, a
whisper the crowd dwarfs) than in selection ($+0.10$ to $+0.14$). Because identity
gaps are within-pair differences, they difference out per-document difficulty and
compare across cohorts better than absolute AP --- though the two tasks still use
different cohorts, so the ratio is approximate. The recoverable
individuality is a real, leakage-free signature that lives far more in what a person
keeps than in what stands out.

\paragraph{Robustness to the embedding model.} Because our headline scorer is the
closed \texttt{text-embedding-3-small}, we re-ran the key contrasts with an
open-weights embedding (\texttt{embeddinggemma-300m}, run locally) on the same pairs,
so the comparison is within-pair. The selectivity identity gap replicates under the
open model: $+0.090$ (95\% CI $[0.061, 0.122]$, positive in 100\% of resamples) versus
$+0.151$ for the OpenAI embedding on that cohort --- about 60\% of the magnitude under
the smaller open model, but the same sign and significance. On within-document
salience, both embeddings agree that the personal gap is a whisper near zero (open
$-0.000$, n.s.; OpenAI $+0.006$, n.s., on that cohort). So the salience--selection
asymmetry --- a large personal signal in selection, essentially none in salience ---
holds under an independent open embedding and is not an artifact of the proprietary
model.

\subsection{How much history? A selectivity scaling law}
\begin{figure}[t]
\centering
\includegraphics[width=0.6\linewidth]{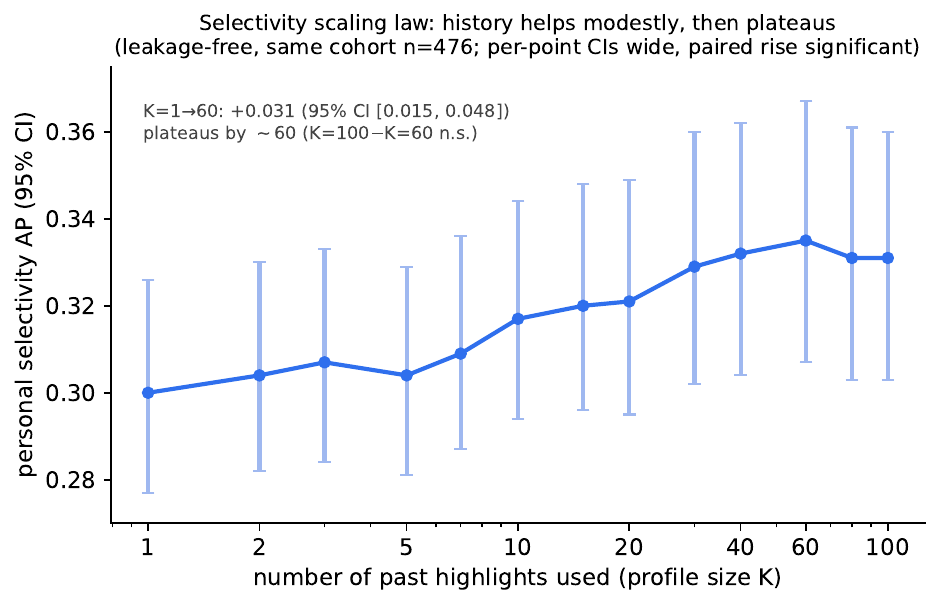}
\caption{Selectivity scaling law (full grid to $K{=}100$, leakage-free, same cohort,
95\% cluster-bootstrap CIs). Per-point CIs are wide, but the paired rise is
significant ($K{=}1\!\to\!60$: $+0.031$, CI $[0.015, 0.048]$); the curve plateaus by
$K\!\sim\!60$ (no gain from 60 to 100). A modest profile suffices.}
\label{fig:scaling}
\end{figure}

Varying the number of past highlights $K$ used to build the personal profile on a
fine grid up to $K{=}100$ (476 pairs, leakage-free, 95\% cluster-bootstrap CIs;
restricted to users with $\geq$100 prior highlights so $K$ can reach 100, hence a
power-user subset), selectivity AP rises from $0.300$ at $K{=}1$ to a plateau
of $\sim$$0.335$ (Figure~\ref{fig:scaling}). The per-point CIs are wide and overlap,
so individual $K$ values are not separable; but the \emph{paired} rise is
significant: $K{=}1\!\to\!60$ is $+0.031$ (95\% CI $[0.015, 0.048]$, positive in
100\% of resamples), with about two-thirds of the gain reached by $K{=}20$ ($+0.021$
vs.\ $K{=}1$). The curve then plateaus: $K{=}100$ minus $K{=}60$ is $-0.004$ (95\% CI
$[-0.013, 0.002]$, not significant). So a person's selectivity is captured by a
modest amount of history --- a few dozen highlights --- after which more does not
help; the crowd, by contrast, is data-hungry, its accuracy still rising at 50
readers.

\subsection{A secondary observation: crowd consensus and document popularity}
\label{sec:decay}
\begin{figure}[t]
\centering
\includegraphics[width=0.58\linewidth]{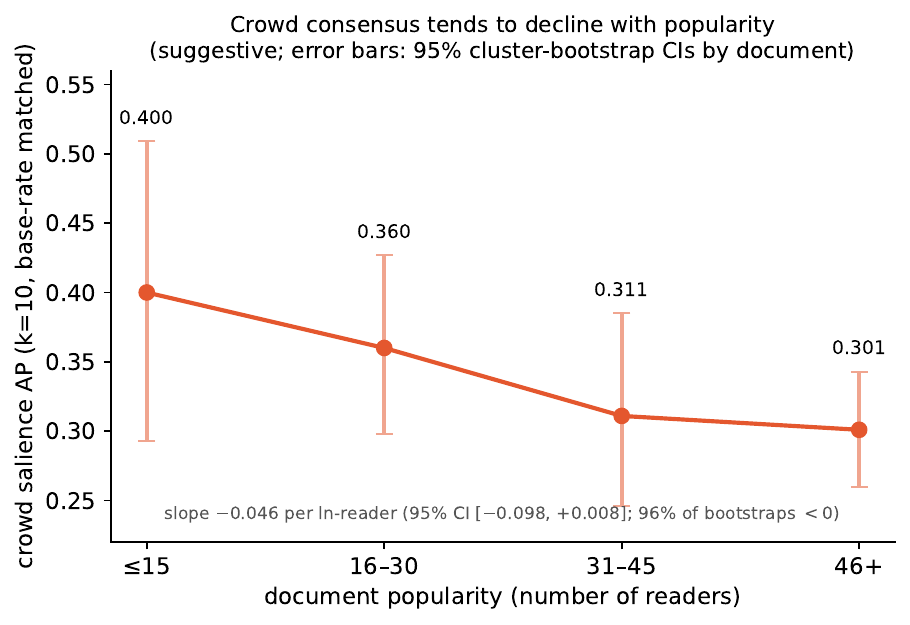}
\caption{Crowd consensus tends to decline with popularity: at a fixed crowd size (10)
and matched base rate, the more readers a document has, the less they agree on what
to highlight. Error bars are 95\% cluster-bootstrap CIs (by document); the trend is
suggestive (slope $-0.046$ per ln-reader, 95\% CI $[-0.098, +0.008]$, negative in
96\% of resamples) and the smallest-popularity bin is sparse.}
\label{fig:decay}
\end{figure}

Finally, an exploratory observation about the crowd itself (not a headline claim): its
internal agreement appears to weaken as a document's audience grows. Holding the crowd
size fixed (10 readers) and matching base rate (mean base rate is flat at
$0.086$--$0.095$ across bins, so the trend is not a base-rate artifact), crowd AP
falls monotonically with the number of readers a document has: $0.400$ ($\leq$15
readers), $0.360$ (16--30), $0.311$ (31--45), $0.301$ (46+) (Figure~\ref{fig:decay}).
The effect is \emph{suggestive} rather than conclusive: a per-document regression
gives a slope of $-0.046$ in AP per natural-log reader (95\% CI $[-0.098, +0.008]$;
negative in 96\% of cluster-bootstraps), and the smallest-popularity bin holds only
$\sim$20 documents, so its interval is wide. To the extent it holds, the more readers
a document attracts, the more heterogeneous they are, and the less they agree on what
to highlight --- a naturalistic, fixed-crowd-size instance of the
annotation-disagreement literature~\cite{aroyo2015}. (This also contrasts with the
individual signal: the crowd is data-hungry, its predictive power still rising at 50
readers, whereas selectivity plateaus by a few dozen highlights.)

\section{Discussion}
Three findings cohere into a single picture. \emph{What} a person highlights is
mostly social: people converge on the same salient passages, an
(information-privileged) crowd is the strongest predictor of an individual, and
neither document structure nor a frontier LLM twin built from the person's history
beats a well-estimated crowd. The recoverable individual signal in salience is, at
most, a whisper --- a small but significant embedding gap of $+0.017$, dwarfed by the
crowd. \emph{Which}
already-salient passage a person selects, however, carries a clear individual
signature --- a leakage-free own-versus-other identity gap, captured by a modest
amount of history (a few dozen highlights for users with rich histories, then it
plateaus). A topic decomposition shows that signature is largely stable thematic
preference: it shrinks $\sim$8$\times$ against a topically-matched peer, though a thin
residual survives even a near-topic-twin. This is not the truism that people read
what interests them: under the same scorer the \emph{same} thematic preference is
about $6$--$8\times$ weaker in within-document salience (where the crowd
rules) than in selection. The individuality is not so much in \emph{what} interests
you, but in \emph{which} of the commonly-interesting things you select.
And the shared layer itself may not be monolithic: crowd
consensus appears to thin as a document's audience grows (a suggestive trend).

\paragraph{Why is salience shared?} We did not test mechanism, but several
non-exclusive explanations are plausible and worth stating. Documents carry strong
rhetorical structure --- topic sentences, thesis statements, conclusions, quotable
formulations --- that pulls most readers toward the same spans. Readers of a given
article also share background knowledge and the cultural framing that makes a passage
``the point.'' And the population that co-reads a document is itself selected (people
who sought out the same article), which compresses variance in what they find worth
marking; this echoes the imitation and shared-knowledge dynamics that stabilize the
relative proportions of tags in collaborative tagging~\cite{golder2006}. These would
all produce high inter-reader agreement on \emph{what} is salient while leaving room
for individuality in \emph{which} salient passage one keeps --- the pattern we
observe. Disentangling them is future work.

For the digital-twin question, the answer is a tempered one. Behavioral trace data
does encode the individual, but on the everyday behavior of highlighting the
individual sits beneath a strong shared layer, and surfaces in selection rather
than in salience. Put sharply: \emph{a digital twin built from highlights predicts
which interesting things you choose, not what you find interesting.} This inverts a
common premise of personalization --- that individuals differ chiefly in what they
find salient. Our evidence suggests the opposite: salience is largely shared, and
individuality emerges in selection among shared salient options. The apparent earlier
result that a frontier twin ``beats the crowd'' was an artifact of two methodological
pitfalls we document: a crowd capped at a few readers, and target-document leakage
into the profile.

\section{Limitations}
The identity control requires shared documents, so results are on moderately
popular content and may not generalize to niche or private reading. The crowd is an
information-privileged baseline: it sees real human marks on the exact target
document, while the twin sees only the user's other-document history; that the
twin loses is therefore not damning, but it does refute the claim that
personalization overturns the crowd. Re-anchoring introduces label noise, and a
small fraction of pairs over-anchor (filtered by a base-rate cap). The profile uses
recent rather than strictly prior documents, so we frame the task as
\emph{recovering} an individual signature, not forecasting. The consensus-decline
trend is only suggestive (its slope CI grazes zero) and, while shown at a fixed
crowd size and base rate, is not fully decomposed from topic and length. A single platform and a single behavior (highlighting) are
studied; richer history aggregation (retrieval, recency) and fine-tuning are
unexplored. Our personal models are two specific extractors (a mean-pooled embedding
and one prompted, proprietary LLM); a stronger personalizer could recover more, so
the within-document ``whisper'' is a lower bound on recoverable signal, not proof of
its absence (we do, however, replicate the asymmetry under an open embedding,
\S\ref{sec:topic}, so it is not specific to the proprietary embedding). On the
salience side, a lead/position prior is in fact a strong baseline (\S\ref{sec:social},
beating semantic centrality and the personal embedding), but the crowd still beats it,
so the crowd's dominance is not merely positional. We decompose the selectivity signal against a topic-matched
peer (\S\ref{sec:topic}) and find it largely thematic; the thin residual that
survives we cannot attribute specifically to non-thematic ``style'' versus
finer-grained topic, which would need structural features or cross-topic transfer. Profile
exclusion is by exact normalized URL; web content is widely syndicated and mirrored
(reposts, aggregators, tracking-parameter variants), so a near-duplicate of the
target at a different URL could still leak. We did not content-deduplicate, so we
cannot rule this out; it would inflate the personal scores in the same direction as
the exact-URL leakage we do correct, and quantifying it (via title or embedding
near-duplicate detection) is left to future work. The scaling-law cohort is power
users ($\geq$100 highlights), so ``a few dozen highlights suffice'' need not hold for
typical users.

\section{Ethics}
The study is retrospective and uses no new recruitment. Because co-readership counts
and highlighting patterns are identifying, the per-pair data are not released; we
publish only aggregate statistics and the analysis code. Copyrighted full text and
frequently sensitive documents are excluded by construction.

\paragraph{Conflict of interest.} The authors are affiliated with Glasp, the platform
whose data are studied. We mitigate the obvious incentive to flatter the product with
a method-matched identity control, leakage-free measurement, released estimator code,
and by foregrounding the unflattering results: a below-chance crowd in the selectivity
task, a within-document personal signal that is a mere whisper, an individuality that
is mostly stable topic preference, and a merely suggestive consensus effect.

\section{Conclusion}
What you highlight is mostly what everyone highlights, and consensus appears to thin
as the audience grows. Beneath the shared layer there is a real individual signature, but
it lives far more in \emph{which} salient passages you select than in what is salient
--- and recovering it requires measuring without leakage and against a crowd that is
not artificially small. Highlights are a partial, selection-shaped window onto the
person, not a clone.

\end{document}